## Journal of Geophysical Research: Space Physics

# Spatial structure and temporal evolution of energetic particle injections in the inner magnetosphere during the 14 July 2013 substorm event

Matina Gkioulidou[1], S. Ohtani[1], D. G. Mitchell[1], A. Y. Ukhorskiy[1], G. D. Reeves[2], D. L. Turner[3], J. W. Gjerloev[1], M. Nosé[4], K. Koga[5], J. V. Rodriguez[6,7], and L. J. Lanzerotti[8]

[1]Johns Hopkins University Applied Physics Laboratory, Laurel, Maryland, USA, [2]Los Alamos National Laboratory, Los Alamos, New Mexico, USA, [3]Space Sciences Department, The Aerospace Corporation, El Segundo, California, USA, [4]Data Analysis Center for Geomagnetism and Space Magnetism, Graduate School of Science, Kyoto University, Kyoto, Japan, [5]Japan Aerospace Exploration Agency, Tsukuba, Japan, [6]Cooperative Institute for Research in Environmental Sciences, University of Colorado Boulder, Boulder, Colorado, USA, [7]National Geophysical Data Center, National Oceanic and Atmospheric Administration, Boulder, Colorado, USA, [8]Center for Solar-Terrestrial Research, New Jersey Institute of Technology, Newark, New Jersey, USA

**Abstract** Recent results by the Van Allen Probes mission showed that the occurrence of energetic ion injections inside geosynchronous orbit could be very frequent throughout the main phase of a geomagnetic storm. Understanding, therefore, the formation and evolution of energetic particle injections is critical in order to quantify their effect in the inner magnetosphere. We present a case study of a substorm event that occurred during a weak storm ($Dst \sim -40$ nT) on 14 July 2013. Van Allen Probe B, inside geosynchronous orbit, observed two energetic proton injections within 10 min, with different dipolarization signatures and duration. The first one is a dispersionless, short-timescale injection pulse accompanied by a sharp dipolarization signature, while the second one is a dispersed, longer-timescale injection pulse accompanied by a gradual dipolarization signature. We combined ground magnetometer data from various stations and in situ particle and magnetic field data from multiple satellites in the inner magnetosphere and near-Earth plasma sheet to determine the spatial extent of these injections, their temporal evolution, and their effects in the inner magnetosphere. Our results indicate that there are different spatial and temporal scales at which injections can occur in the inner magnetosphere and depict the necessity of multipoint observations of both particle and magnetic field data in order to determine these scales.

## 1. Introduction

Energetic particle injections have been extensively studied at geosynchronous orbit leading to the model of the "injection boundary." According to that model, accelerated plasma occupies a region that is extended in magnetic local time (MLT) and lies tailward of a boundary, which represents an initial condition in the inner magnetosphere [e.g., *McIlwain*, 1974; *Mauk and Meng*, 1983; *Birn et al.*, 1997; *Thomsen et al.*, 2001]. However, this concept was based solely on particle observations, and the goal was to explain the various energy dispersion signatures of injections that could emerge from energy-dependent particle drifts.

In the magnetotail and near-Earth plasma sheet, energetic particle injections have been correlated with transient, localized in MLT dipolarization fronts, fast flows, and enhanced electric fields [e.g., *Baumjohann et al.*, 1990; *Nakamura et al.*, 2002; *Runov et al.*, 2009; *Gabrielse et al.*, 2014]. The structures that feature all the above characteristics have been interpreted as "plasma bubbles," that is, depleted flux tubes with entropy lower than that of the neighboring ones. These depleted flux tubes move earthward due to interchange instability [*Pontius and Wolf*, 1990]. Their spatial and temporal evolution as they propagate earthward into the transition region from the taillike to dipolar magnetic field configuration, and how if at all they fit the injection boundary picture in the inner magnetosphere, is still under debate.

Recent results from the Van Allen Probes mission showed that the occurrence of energetic ion injections inside geosynchronous orbit can be very frequent throughout the main phase of a geomagnetic storm and indicated that the contribution of such injections to the pressure buildup could be substantial [*Gkioulidou et al.*, 2014; *Yu et al.*, 2014]. Also, various wave modes and electrostatic structures have been observed in the vicinity of injection fronts inside geosynchronous orbit [*Mozer et al.*, 2013; *Chaston et al.*, 2014; *Malaspina* *et al.*, 2014]. These waves could potentially accelerate the electron seed population up to MeV energies and contribute to the enhancement of the radiation belts [e.g., *Reeves et al.*, 2013; *Thorne et al.*, 2013].

In order to quantitatively investigate the effect of particle injections into the inner magnetosphere, it is critical to determine their spatial extent and temporal evolution. In this paper we investigate the spatial and temporal characteristics of two injections Van Allen Probe B observed during a substorm event that occurred on 14 July 2013. For this purpose we use multipoint observations from spacecraft in the inner magnetosphere and near-Earth plasma sheet, as well as ground-based magnetometers. In section 2.1 we give an overview of the event, in section 2.2 we describe the in situ particle and magnetic field observations, in section 2.3 we describe the ground-based magnetometer observations, and in section 3 we discuss the possible interpretations of the event based on the data synthesis.

## 2. Multipoint Observations During the Substorm Injection Event

### 2.1. Overview of the 14 July 2013 Injection Event

On 14 July 2013, Van Allen Probe B (hereafter referred to as RBSP-B) observed two energetic proton injection events with different timescales and magnetic field signatures approximately 10 min apart (at ~ 11:21 and ~ 11:32 UT). During that time interval RBSP-B was located at $L \sim 5.5$ to $L \sim 5$ and MLT $\sim 21$ h. Figure 1 is an overview of the event. During the time interval of interest, 11:00–12:00 UT, we show (a and b) the solar wind conditions, (c and d) geomagnetic indices, (e) $Wp$ index related to the wave power of low-latitude Pi2 pulsations, (f–g) 90° pitch angle proton intensities from Radiation Belt Storm Probes Ion Composition Experiment (RBSPICE) [*Mitchell et al.*, 2013], and (h) the $B_z$ component and total magnitude of the magnetic field in SM coordinates from the Electric and Magnetic Field Instrument Suite and Integrated Science (EMFISIS) [*Kletzing et al.*, 2013], both instruments on board RBSP-B.

The interplanetary magnetic field (IMF) remains southward throughout the interval, and *SYM-H* index is $\sim -40$ nT until ~11:30 UT; thus, a small geomagnetic storm is in progress [*Gonzalez et al.*, 1994]. From ~ 11:15 UT to ~ 11:40 UT the $Wp$ index is elevated, indicating low-latitude Pi2 pulsations [*Nosé et al.*, 2012], which are closely related to substorm activity. After ~11:32 UT the *SYM-H* index gradually increases. Since there is no dynamic pressure enhancement at the time, the *SYM-H* index increase can be attributed to the magnetic field perturbations due to the field-aligned currents after the development of the substorm current wedge [*McPherron et al.*, 1973]. The *AL* index shows a significant decrease a few minutes later, at ~ 11:36 UT. Therefore, a substorm current wedge was formed after 11:30 UT, yet the exact substorm onset time is hard to define by the *AL* index decrease, since preexisting geomagnetic activity (*AL* index is $\sim -200$ to $-400$ nT prior to 11:36 UT) can mask the effect of the auroral electrojet at the early stages of the substorm expansion phase.

The Van Allen Probes mission consists of two identically instrumented spacecraft (RBSP-A and RBSP-B) in nearly identical ~ 9 h orbits, with perigee of ~ 600 km altitude, apogee of 5.8 $R_E$, and inclination of 10° [*Mauk et al.*, 2012]. As mentioned above, the RBSPICE instrument on RBSP-B observed two energetic proton injection events. Figure 1e shows the energy spectrogram for 45–600 keV protons, and Figure 1f shows proton intensities versus time for energies 24–327 keV. The two dotted lines in Figure 1 indicate the time when the intensity of the 268 keV protons starts rising for each injection event. The first injection was observed at ~ 11:21 UT and was dispersionless; that is, the intensities of protons of all energies between 81 and 268 keV sharply increase about an order of magnitude simultaneously. The intensities of 127 keV–268 keV protons rise and drop to their preinjection values within 3 min or less depending on the energy (from ~ 11:21 to ~ 11:24 UT), while 81–99 keV proton intensities stay elevated longer. The magnetic field sharply dipolarizes at 11:21:30 UT; both the $B_z$ component and the total magnetic field magnitude increase by ~ 30 nT in a few seconds and remain elevated until ~ 11:33 UT. Note that intensities of protons of energies below 67 keV slightly increase right before the dipolarization and then decrease when the more energetic proton intensities increased. Certain properties of the second injection event are quite different from the first one. During the second injection the intensity enhancements exhibit energy dispersion, with higher-energy proton intensities rising earlier than the lower energy ones (~1.5 min delay between the rise of 81 and 327 keV protons). This means that the spacecraft is not inside the particle acceleration region at that time, and thus, the higher-energy protons arrive to the spacecraft location faster than the lower energy ones due to westward energy-dependent magnetic drift. Protons of energies lower than 67 keV behave similarly to the first injection with their intensities slightly increasing before and decreasing during the enhancement of the higher-energy proton intensities. This energetic proton injection event lasts roughly from 11:32 UT, when intensity of 327 keV protons starts increasing, to ~ 11:44 UT when intensities of protons from 147–220 keV drop significantly and very close to their preinjection values. Intensities of protons of 81–121 keV also decrease but remain elevated with respect to preinjection values. Therefore, unlike the first injection, which was of much shorter timescale, the second injection pulse lasted for almost 12 min. However, during these 12 min there are embedded several shorter-timescale intensity enhancements. There is no sharp dipolarization observed during this event but instead a gradual increase of the $B_z$ component and total magnetic field from ~ 11:34 UT to ~ 11:44 UT.

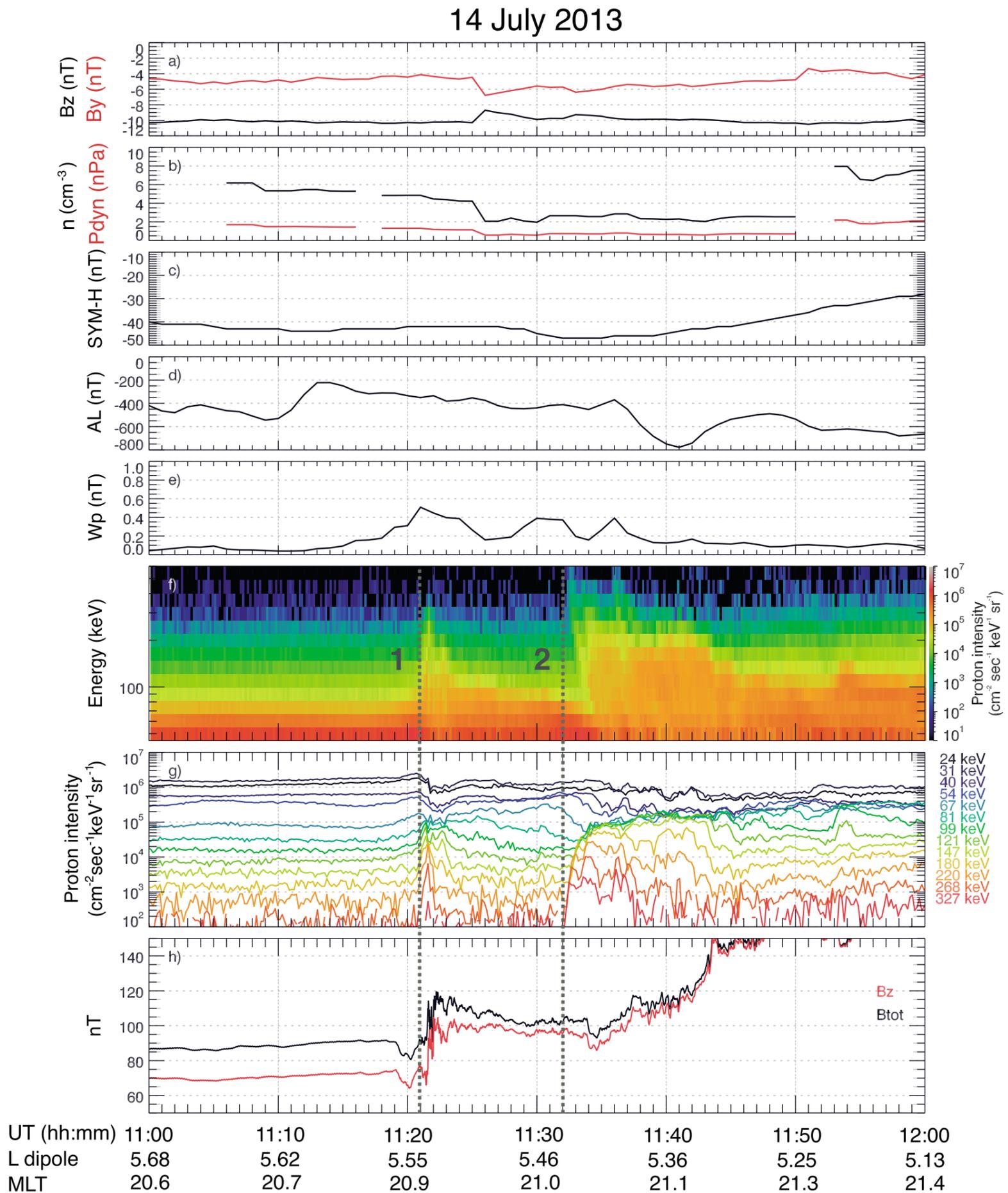

**Figure 1.** (a) Interplanetary magnetic field (IMF) $B_z$ and $B_y$ components, (b) solar wind number density and dynamic pressure, (c) *SYM-H* index, (d) *AL* index, (e) *Wp* index, (f and g) RBSPICE proton intensities, and (h) EMFISIS magnetic field component $B_z$ and total magnetic field $B_{tot}$. Solar wind data are acquired from ACE and Wind spacecraft and have been shifted to the bow shock nose.

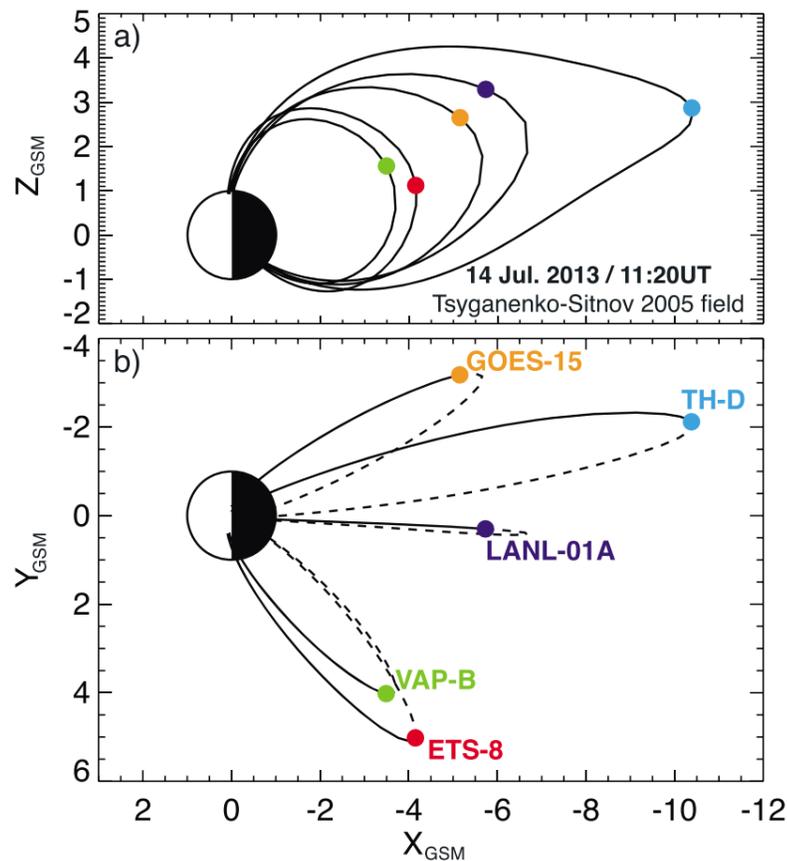

**Figure 2.** Satellite position in GSM coordinates along with TS05 magnetic field lines projected onto (a) the noon-midnight meridional plane and (b) the equatorial plane.

From the description above, it is clear that RBSP-B, located inside geosynchronous orbit, observed two injection events with distinctly different properties. The first event is a dispersionless, short-timescale (~3 min) energetic proton injection accompanied by sharp dipolarization of the magnetic field. The second event is a dispersed energetic proton injection exhibiting ~1.5 min delay in the starting times of intensity increase between the highest and lowest energy channels, lasts significantly longer than the first one (~12 min), and is accompanied by a gradual dipolarization of the magnetic field. Since the two events were observed within 10 min, while the satellite essentially remains in the same region in space, important questions arise: Is the timescale of the intensity enhancement due to energetic particle injections, as observed by a single spacecraft, indicative of the spatial extent or temporal evolution of the injection? Is the magnetic field reconfiguration associated with a particle injection inside geosynchronous orbit related to its duration? Finally, how do energetic particle injections that penetrate inside geosynchronous orbit change the plasma properties, such as density, temperature, and pressure in the inner magnetosphere? Are the changes similar to those during energetic particle injections in the tail?

In order to differentiate between the spatial and temporal components of the injections in the inner magnetosphere, it is necessary that we combine multipoint observations. In the following sections we use (i) in situ particle and magnetic field observations from four spacecraft located in the inner magnetosphere and near-Earth plasma sheet and (ii) ground-based magnetic field data from several stations to investigate the inner magnetosphere conditions as well as the global current systems throughout the interval when the two injections were observed by RBSP-B. Our goal is to explore the possible scenarios that would result in the different properties characterizing these injections.

## 2.2. Multipoint In Situ Particle and Magnetic Field Observations

Between 11:20 and 11:30 UT there were several spacecraft scanning the nightside magnetosphere. Figure 2 shows the position of these spacecraft at 11:20 UT in GSM coordinates projected onto the (a) noon-midnight meridional and (b) equatorial planes, along with magnetic field lines obtained by the Tsyganenko-Sitnov 2005 magnetic field model [*Tsyganenko and Sitnov*, 2005]. Geosynchronous satellites GOES 15, Los Alamos National Laboratory (LANL)-01A, and ETS-8 are located postmidnight (MLT ~ 2 h), midnight, and premidnight (MLT ~ 21 h), respectively, Time History of Events and Macroscale Interactions during Substorms (THEMIS)-D (TH-D), at $r \sim 11\, R_E$, is very close to midnight (MLT ~ 1 h), and RBSP-B is located premidnight at the same MLT as ETS-8 (MLT ~ 21 h) but at $r \sim 5.5\, R_E$. The ETS-8 satellite, launched on 18 December 2006 by the Japan Aerospace Exploration Agency, carries a triaxial fluxgate magnetometer (for more details, read *Nosé et al.* [2014]).

Figure 3 shows proton intensities from (a) the RBSPICE instrument on RBSP-B, (b) the Synchronous Orbit Particle Analyzer (SOPA) detector on LANL-01A, and (c) the Magnetospheric Proton Detector (MAGPD) on GOES 15. As we have already discussed in section 2.1, RBSP-B at MLT ~ 21 h and $r \sim 5.5\, R_E$ observed two major injections, a dispersionless one at 11:21 UT and a dispersed one at 11:32 UT. Geosynchronous satellite

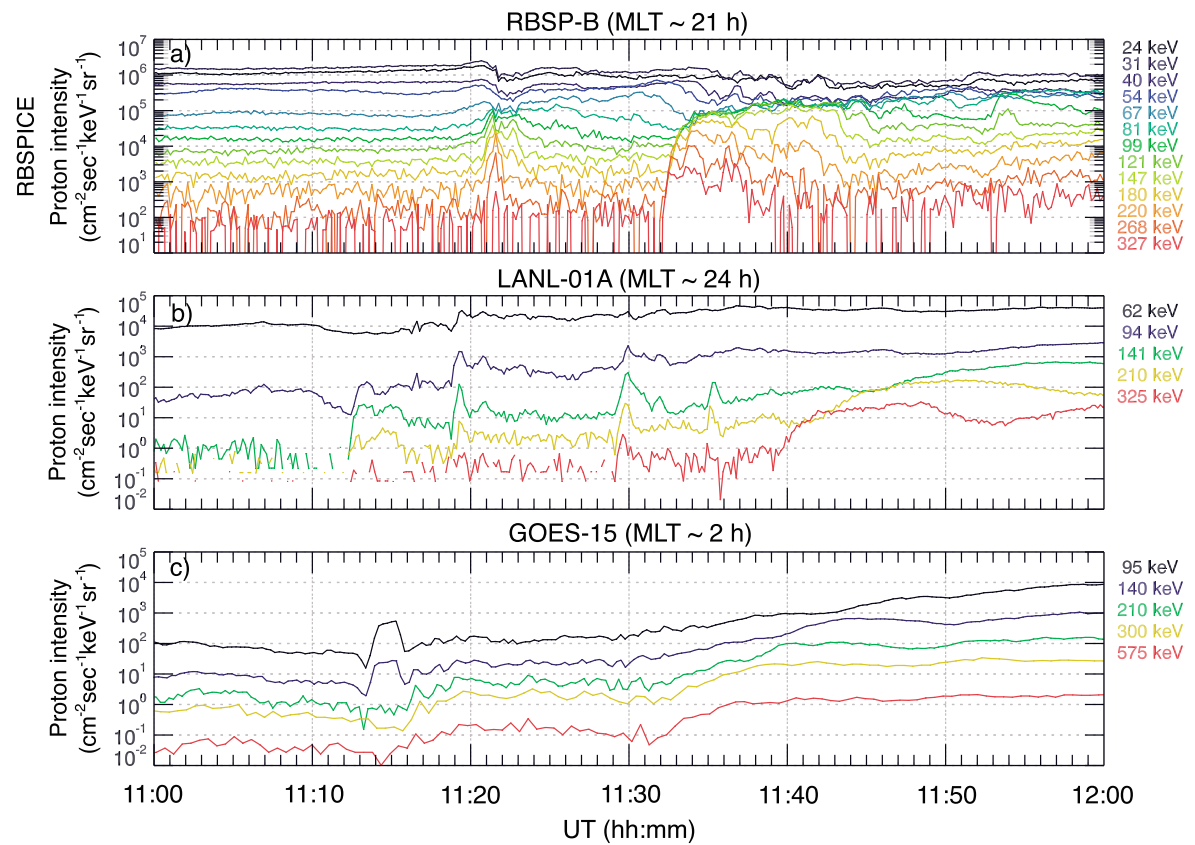

**Figure 3.** (a) RBSP-B/RBSPICE, (b) LANL-01A/SOPA, and (c) GOES 15/MAGPD omnidirectional proton intensities.

LANL-01A at MLT ~ 24 h observed three major dispersionless enhancements of the high-energy proton intensities at 11:12 UT, 11:19 UT, and 11:29 UT. Around ~ 11:39 UT there is the dispersed signature of a drift echo from protons that have already drifted around the Earth once. Geosynchronous satellite GOES 15 at MLT ~ 2 h observed only one energetic particle injection at 11:13 UT in the 95 keV and 140 keV channels, while the 210–575 keV channels increase at ~11:16 UT. There is no other injection signature for the rest of the time interval, only a slow rise of the intensities starting at ~11:32 UT.

Figure 4 shows electron intensities observed by the same spacecraft as in Figure 3. The Magnetic Electron Ion Spectrometer (MagEIS) [*Blake et al.*, 2013] instrument on RBSP-B observed only one dispersionless energetic electron injection at ~ 11:21:30 UT but no injection at 11:32 UT. There is a simple explanation for that: since RBSP-B is located premidnight, it would miss any energetic electron injection that occurred east of the spacecraft (as it is inferred by the energy dispersion in the proton injection) because electrons, as opposed to protons, would drift farther eastward. LANL-01A, located at midnight, observed a reduction in the electron intensities starting at ~ 11:10 UT and a sudden enhancement at ~ 11:15 UT. The intensities stay approximately at the same level (smaller enhancements are embedded) until ~ 11:30 UT when they start gradually increasing after a very short reduction (no signature of impulsive injection) until 11:34 UT. They remain elevated until 12:00 UT. The Magnetospheric Electron Detector (MAGED) on GOES 15, 2 h eastward of LANL-01A, observes very similar electron intensity profiles but with ~1 min delay. Unlike LANL-01A, GOES 15 also observes ~ 1 min period modulations of the electron intensities between 11:29 UT and 11:32 UT.

Figure 5 shows proton and electron data (a and b) along with the three components of the magnetic field (c) and velocity vector (d–f) of the TH-D spacecraft. From 11:15 UT to 11:38 UT TH-D observes very strong fluctuations of the $B_x$ and $B_y$ components of the magnetic field, while $B_z$ remains low. The fluctuations from 11:15 UT to 11:20 UT are indicative of possible flapping motion of the current sheet [e.g., *Sergeev et al.*, 2003; *Sitnov et al.*, 2014]. During the same time interval very strong flows are observed, tailward (up to ~ −300 km/s) between 11:15 UT and 11:19 UT and earthward (up to ~ 800 km/s) between 11:19 UT and 11:29 UT. At ~ 11:25 UT $B_x$ starts decreasing, while $B_z$ starts increasing. After 11:29 UT, the $B_z$ component becomes the dominant one, thus the spacecraft is in the current sheet. The magnetic field keeps dipolarizing until 11:36 UT and remains constant at ~20 nT thereafter.

In Figure 6 we show the magnetic field components measured by magnetometers on RBSP-B, GOES 15, and ETS-8 spacecraft in *VDH* coordinates (*V*: positive tailward, *D*: positive eastward, and *H*: positive northward) and the TH-D magnetic field components in GSM coordinates. Below we summarize all the in situ particle and magnetic field observations and discuss the possible implications:

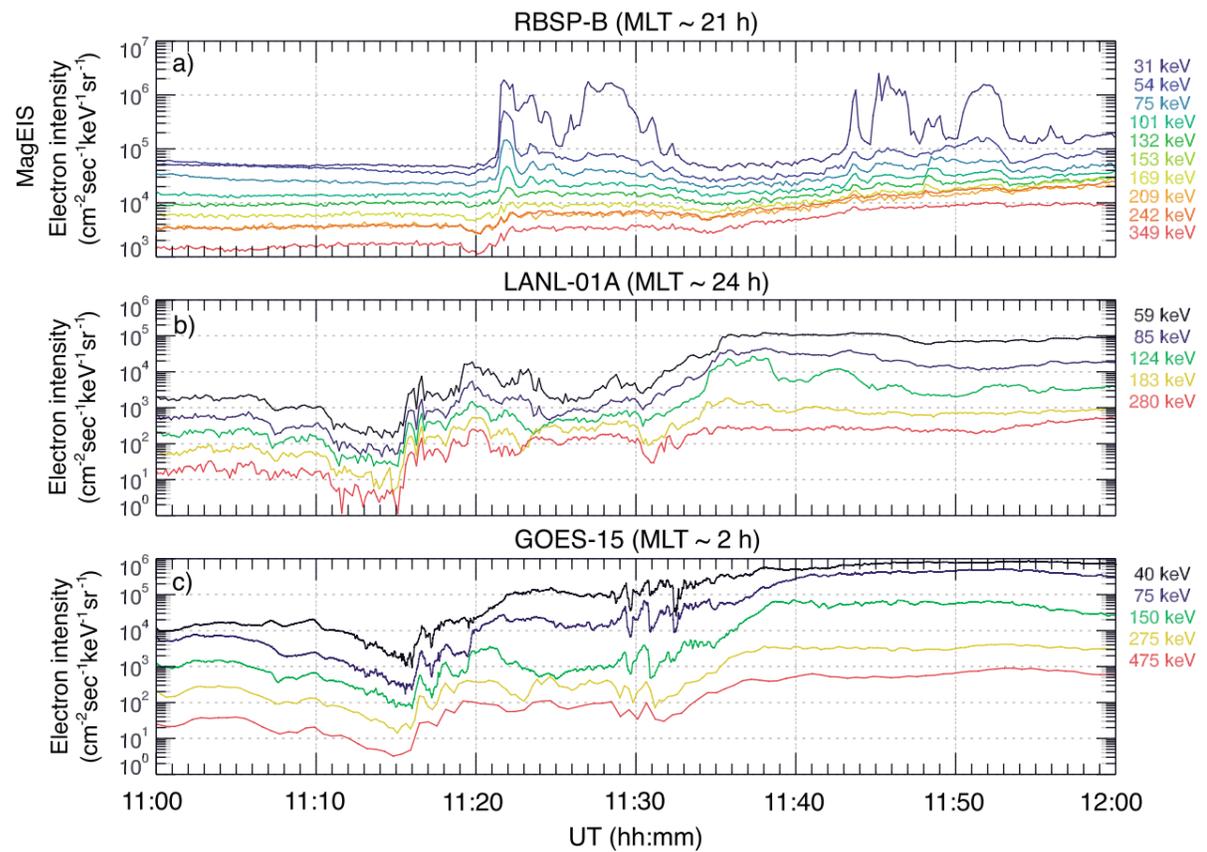

**Figure 4.** (a) RBSP-B/MagEIS, (b) LANL-01A/SOPA, and (c) GOES 15/MAGED omnidirectional electron intensities.

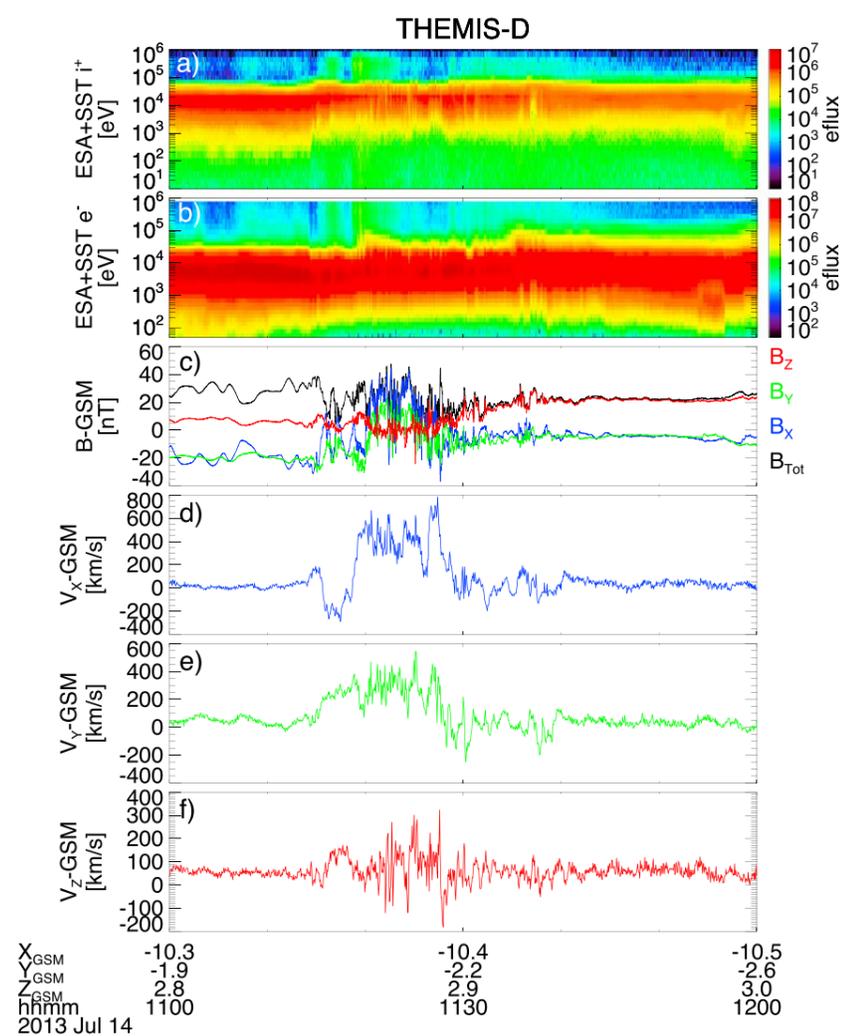

**Figure 5.** TH-D (a) combined ESA and SST ion intensities; (b) combined ESA and SST electron intensities; (c) magnetic field components $B_x$, $B_y$, $B_z$, total magnetic field $B_{tot}$; and velocity components (d) $V_x$, (e) $V_y$, and (f) $V_z$.

1. GOES 15 ($r \sim 6.6\,R_E$, MLT $\sim 2$ h). The $V$ component of the GOES 15 magnetic field (yellow line in Figure 6a), which is the prominent one at least until $\sim 11{:}35$ UT, increases in magnitude from 11:07 UT to 11:13 UT when it starts decreasing. Also, electron (Figure 4c) and proton (Figure 3c) intensities start decreasing at $\sim 11{:}07$ UT, and while low-energy protons recover at $\sim 11{:}13$ UT exhibiting an injection signature, high-energy protons and electrons recover at $\sim 11{:}16$ UT. In fact, electrons do not reach the intensity levels prior to the decrease until $\sim 11{:}20$ UT. The above magnetic field and particle observations can be attributed to the thinning ($V$ component magnitude increase, intensities decrease) prior to the thickening ($V$ component magnitude decrease, intensities increase) of the near-Earth plasma sheet. Similar results were reported by *Angelopoulos et al.* [1996] during a bursty bulk flow event, where a geosynchronous satellite observed thinning and recovery of the plasma sheet with proton intensities recovering earlier and exhibiting a much sharper increase than the electron ones.

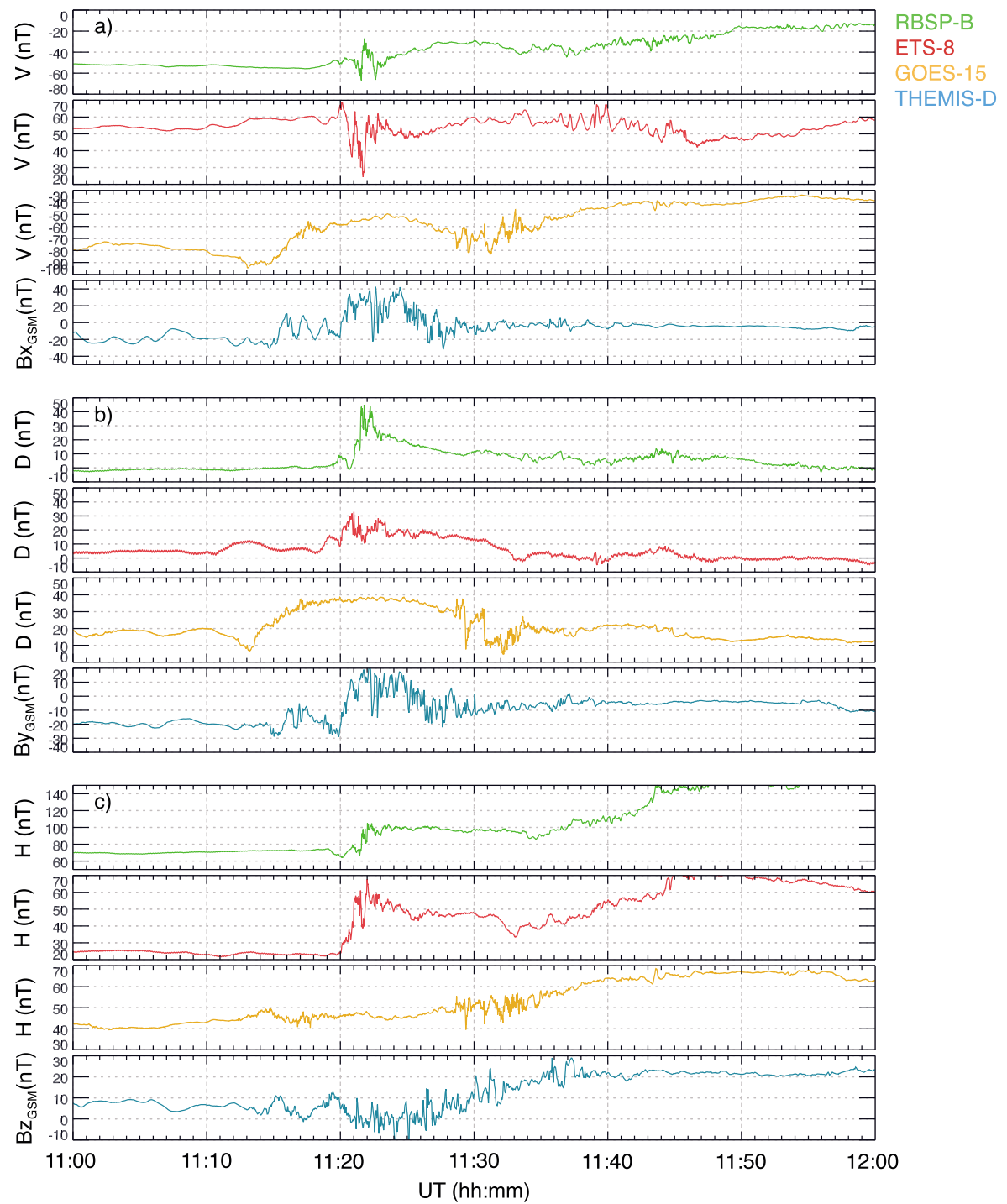

**Figure 6.** (a) $V$ component of RBSP-B/EMFISIS, ETS-8, and GOES 15 magnetic field and $B_x$ component of TH-D magnetic field in GSM coordinates; (b) $D$ component of RBSP-B/EMFISIS, ETS-8, and GOES 15 magnetic field and $B_y$ component of TH-D magnetic field in GSM coordinates; and (c) $H$ component of RBSP-B/EMFISIS, ETS-8, and GOES 15 magnetic field and $B_z$ component of TH-D magnetic field in GSM coordinates.

2. TH-D ($r \sim 11~R_E$, MLT $\sim 1$ h). Around the same time that the beginning of the plasma sheet thickening is observed by GOES 15 (~11:13 UT–11:15 UT), all the magnetic field components of TH-D (blue lines) start fluctuating accompanied by fast flows until 11:29 UT. Although the fast flows are tailward between 11:15 UT and 11:19 UT, after 11:19 UT we have the initiation of a bursty bulk flow (BBF) event. Usually, BBFs are accompanied by a dipolarization signature [e.g., *Angelopoulos et al.*, 1992]. However, since the spacecraft is not always in the center of the current sheet, as is evident by the very strong fluctuations in $B_x$ component and the low values of $B_z$, it would be difficult to observe such signature. Nonetheless, GOES 15 farther earthward and only 1 h eastward of TH-D observes the thickening of the plasma sheet.

3. LANL-01A ($r \sim 6.6~R_E$, MLT $\sim 24$ h). The timing of the first proton injection that LANL-01A sees at ~11:12 UT (Figure 3b) coincides approximately with the timing of the proton injection at GOES 15 (Figure 3c). Also, similar to GOES 15, electron intensities at LANL-01A (Figure 4b) decrease from 11:07 UT to 11:15 UT when they recover indicating the transition from thinning to thickening of the plasma sheet. Interestingly, after 11:16 UT, that is, after the onset of the plasma sheet thickening, both electron and proton spectra of the two geosynchronous satellites exhibit differences. For example, none of the two proton dispersionless injections that LANL-01A sees after 11:16 UT are detected by GOES 15, which is located 2 h eastward. On the other hand, only GOES 15 observes the ~1 min modulation of the electron intensities around 11:30 UT. The modulation of the electron intensities at GOES 15 also coincides with fluctuations in the $V$ and $H$ magnetic field components with the same period.

4. RBSP-B ($r \sim 5.5\ R_E$, MLT ~ 21 h) and ETS-8 ($r \sim 6.6\ R_E$, MLT ~ 21 h). As mentioned earlier, RBSP-B observes a dispersionless, short-timescale injection in both proton (Figure 3a) and electron (Figure 4a) spectra at 11:21 UT and a dispersed, longer-timescale proton injection at 11:32:30 UT. Although the first injection is accompanied by a sharp dipolarization of the magnetic field starting at 11:21:30 UT (green lines), the second one is accompanied by a gradual one, starting at 11:34:30 UT. Magnetic field measured by ETS-8 (red lines), located at the same MLT as RBSP-B but ~ 1 $R_E$ tailward, shows the same signatures as RBSP-B but starting 1.5 min earlier (sharp dipolarization starts at 11:20 UT and the gradual one at 11:33 UT). The sharp dipolarization has the characteristics of a dipolarization front penetrating into the inner magnetosphere (more discussion follows in section 3.2). With the two spacecraft being located ~ 1 $R_E$ apart, 1.5 min delay infers an ~ 70 km/s earthward propagation speed of the dipolarization, which is smaller than the 180 km/s earthward propagation dipolarization speed reported by *Ohtani* [1998] at $r \sim 7.3\ R_E$ and larger than previously reported average radial propagation speed (24 km/s) of particle injection signatures inside geosynchronous orbit [*Reeves*, 1996]. It is noteworthy that the gradual dipolarization is also observed by GOES 15 (starting at ~ 11:33 UT, when $V$ and $H$ components start decreasing and increasing, respectively) as well as by TH-D (starting at ~ 1:25 UT, when $B_x$ and $B_z$ components start decreasing and increasing, respectively). From the timing described above, it is evident that the gradual dipolarization, unlike the sharp one, is a signature that is observed by all spacecraft at different local times and it propagates earthward.

### 2.3. Ground Magnetometer Data

In order to put the satellite measurements in a global context, in this section we analyze data from various ground magnetometer stations.

Pi2 pulsations are periodic, ultralow-frequency pulsations that can be generated by the onset of field-aligned currents (FACs) and fast-mode waves in the nightside magnetosphere associated with the substorm onset. Figure 7 shows band-pass-filtered (between 40 and 150 s) magnetic field data from several ground stations. Premidnight stations MMB (MLT = 18 h, MLAT = 37°) and KAK (MLT = 21 h, MLAT = 29°) see Pi2 activity from 11:20 UT to 11:25 UT. Midnight stations SHU (MLT = 24 h, MLAT = 53°) and HON (MLT = 24 h, MLAT = 21°) see two major Pi2 onsets, one at ~ 11:15 UT and one at ~ 11:28 UT. Finally, the postmidnight station VIC (MLT = 2 h, MLAT = 54°) also sees two onsets, one at ~ 11:15 UT and one at ~ 11:28 UT.

During the expansion phase of magnetospheric substorms, tail current disruption leads to the formation of a FAC pair: downward FAC on the eastward edge of the current disruption region, where tail current diverts down field lines into the ionosphere, and upward FAC on the westward edge, where current is supplied by the ionosphere up field lines. The westward auroral electrojet closes the current loop in the ionosphere. The current system described above is referred to as the substorm current wedge (SCW) [*McPherron et al.*, 1973]. The upward and downward FAC of the SCW as well as the westward auroral electrojet cause perturbations in the eastward ($B_E$) and northward ($B_N$) magnetic field components measured at middle- and high-latitude stations. More specifically, midlatitude stations observe positive perturbation of the $B_N$ component when located inside the current wedge due to both upward and downward FAC and positive (negative) perturbation of the $B_E$ component when located on the westward (eastward) edge of the current wedge due to upward (downward) FAC. High-latitude stations inside the current wedge would observe negative $B_N$ perturbation (negative bay) due to the westward auroral electrojet. Similar Region 1 sense FAC pair develops on the east (downward FAC) and west (upward FAC) sides of a low-entropy plasma bubble [*Yang et al.*, 2012]. The magnitude of these FACs, and as a consequence the perturbations seen on the ground, depends on the entropy depletion in the bubble and thus could be smaller than of those associated with a large-scale substorm current wedge.

  

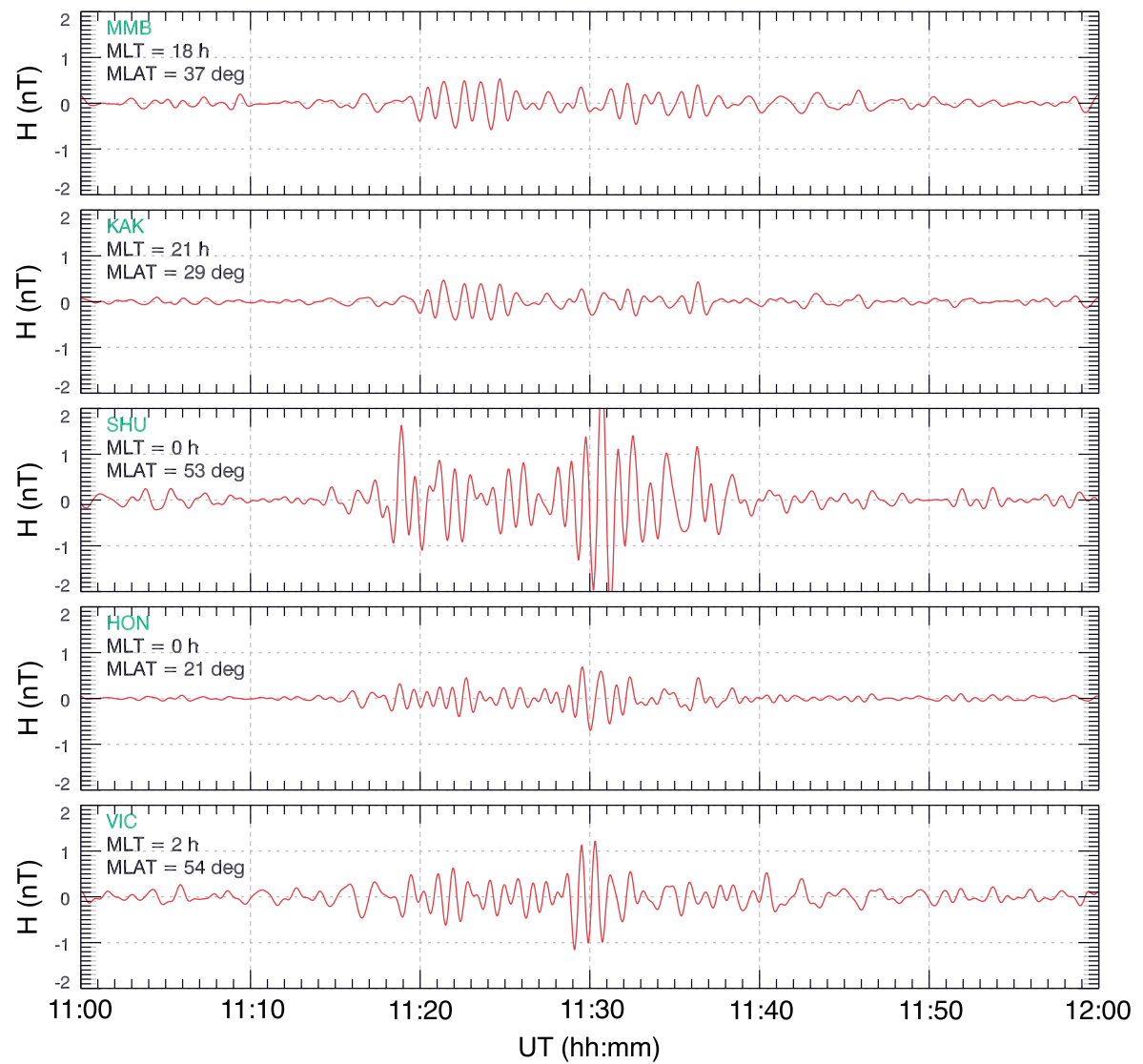

**Figure 7.** Band-pass-filtered (between 40 and 150 s) northward magnetic field component from MMB, KAK, SHU, HON, and VIC ground magnetometer stations.

Figure 8 shows 1 min resolution northward ($B_N$ in red) and eastward ($B_E$ in black) magnetic field components from middle-, low-, and high-latitude stations at various local times, after the values at 11:00 UT have been subtracted. We examine the magnetic field perturbations in the premidnight and postmidnight sectors. Low-latitude KAK and midlatitude MGD stations both at ~ MLT = 21 h see a slight increase of $B_E$ and $B_N$ components at ~ 11:20 UT. The positive $B_N$ perturbation being larger than the $B_E$ one indicates that these stations are duskward of the center and near the upward FAC of a current wedge-like formation. Similarly, BRW, at MLT ~ 23 h but higher latitude (71°), observes a small negative $B_N$ perturbation at ~ 11:20 UT. Hence, BRW is also inside a current wedge-like formation. Note that these small perturbations in the premidnight stations last only until ~ 11:30 UT (the beginning and end of the perturbations are marked by magenta dashed lines in Figures 8a–8c). At the same time that the small perturbations at the premidnight stations are observed (~ 11:20 UT), RBSP-B above the magnetic equator (negative $V$ component) observes a positive $D$ component perturbation. Such perturbation can be attributed to an upward FAC tailward of RBSP-B. Also, as mentioned above, there is an onset of Pi2 pulsations observed by premidnight stations (KAK and MMB) at ~ 11:20 UT, and the main activity lasts until ~ 11:25 UT. The timing of the ground and in situ magnetic field perturbations premidnight coincides with the timing of the first energetic particle injection and dipolarization front observed by RBSP-B. Therefore, from the combination of all the data at hand, we can infer that the ground magnetic field perturbations at ~11:20 UT are a result of an upward FAC forming close to the westward edge of low-entropy bubble. At ~ 11:30 UT, a second, much stronger positive $B_E$ perturbation, accompanied by a negative $B_N$ one, is observed at the KAK and MGD stations. Also, a very strong negative $B_N$ perturbation is observed at the high-latitude BRW station. These perturbations are consistent with the formation of a large-scale SCW, whose westward edge is at ~ 21 h in MLT (KAK and MGD perturbations infer that the stations are outside that current wedge). Interestingly, midnight and postmidnight stations observe different magnetic field perturbations. The perturbations at these local times appear to start slightly earlier. At ~ 11:18 UT, HON (low-latitude) and CMO (high-latitude) midnight stations observe the beginning of positive $B_N$ and negative $B_N$ perturbations, respectively, while VIC at MLT ~ 2 h sees a negative $B_E$ perturbation. All the above perturbations intensify after ~ 11:30 UT. These perturbations imply that stations HON and CMO are inside, and station VIC is at the eastward edge (downward FAC) and outside a current wedge that starts forming at 11:18 UT (cyan line in Figures 8d–8f) but significantly intensifies at ~ 11:30 UT. Also, in contrast to premidnight stations, there are two major Pi2 pulsation onsets, one at ~ 11:18 UT and one at 11:30 UT with activity continuing in between, observed by midnight (SHU and HON) and postmidnight stations (VIC).

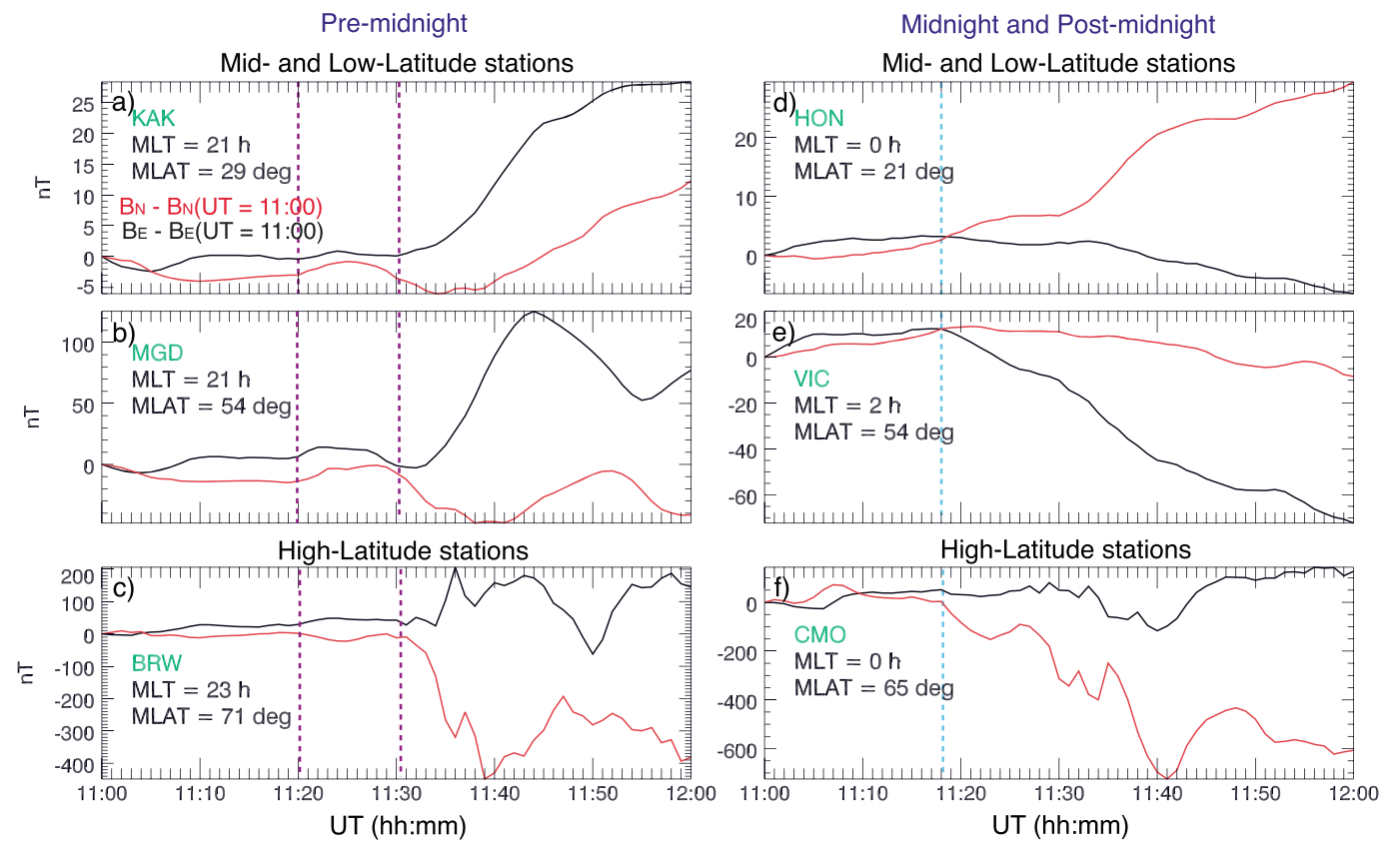

**Figure 8.** One minute resolution northward ($B_N$) and eastward ($B_E$) magnetic field components from KAK, HON, and VIC midlatitude stations at various local times and BRW high-latitude station, after the values at 11:00 UT have been subtracted.

In the following section we combine in situ and ground-based observations and discuss possible interpretations of the event.

## 3. Data Synthesis and Discussion

### 3.1. Correlation Between Energetic Particle Injections and Substorm-Type Current Wedge Formation

Ground magnetometer observations reveal (i) the beginning of a negative bay development at ~ 11:18 UT at midnight and postmidnight stations (from 0 to 2 h); (ii) small perturbations and Pi2 pulsations at ~11:20 UT, at premidnight stations (from ~ 21 h to ~ 23 h), that coincide with the injection observed at RBSP-B (also located at ~ 21 h at that time); and (iii) the formation of a large-scale SCW at ~ 11:30 UT, at both premidnight and postmidnight stations, extending ~ 5 h in MLT (from ~ 21 h to ~ 2 h).

Next, we look into the in situ magnetic field and particle data from the various satellites scanning the inner magnetosphere. The onset of the thickening of the near-Earth plasma sheet at ~ 11:15 UT, as inferred from the electron intensity enhancements observed by LANL-01A and GOES 15 satellites located at midnight and MLT ~ 2 h, respectively, coincides with the beginning of a negative bay forming from ~ midnight to ~ 2 h, which indicates a current disruption region in the magnetosphere at those MLTs. Note that the RBSP-B at MLT ~ 21 h does not observe an injection, proton or electron, at that time. Although that could be interpreted as a nonpropagation of the injection inside geosynchronous, the fact that ETS-8, also at geosynchronous but at the same MLT as RBSP-B, does not observe any dipolarizations at that time means that the injection, and thus the current disruption region, did not extend to that MLT. The small perturbations and Pi2 pulsations at ~ 11:20 UT observed at the premidnight stations (~21 h–23 h) coincide with the dispersionless proton and electron injection at RBSP-B at ∼ 21 h. As we discuss in sections 3.2 and 3.3 below, these observations indicate the existence of a low-entropy bubble extending ∼ 2 h in MLT. Finally, the dispersed proton injection that RBSP-B observes at 11:32 UT coincides with the formation of the large-scale SCW extending ∼ 5 h in MLT. KAK and MGD stations, at the same MLT as RBSP-B, are outside the SCW and on the westward edge, in the upward field-aligned current region. Since RBSP-B is outside the current disruption region, that is the main injection region, it completely missed the electrons that have drifted eastward of the spacecraft. Surprisingly, LANL-01A, which is inside the SCW after 11:30 UT, does not observe a prolonged dispersionless injection in either the proton or electron spectra, as one would expect. It observes only a short-timescale enhancement of proton intensities at ∼ 11:29 UT and a gradual enhancement of the electron intensities starting at 11:30 UT. Similarly, GOES 15, on the eastward edge of the SCW, does not observe a dispersed electron injection, as would be expected from the electron eastward drift. We discuss the implications of these observations in section 3.3.

From the combination of the in situ and ground magnetometer data it is evident that there are different current systems associated with the injections observed by the various satellites in the inner magnetosphere during this particular substorm event:

1. At ∼ 11:18 UT. A negative bay and a downward FAC-associated perturbation start developing at the high-latitude midnight CMO and midlatitude postmidnight (∼2 h) VIC stations, respectively, a few minutes after the thickening of the plasma sheet is observed by geosynchronous LANL-01A (midnight) and GOES 15 (∼2 h) at ∼ 11:15 UT. Ground magnetometer perturbations keep intensifying at the midnight and postmidnight stations.
2. At ∼11:20 UT. A weak, localized current system develops premidnight (∼21–23 h) coinciding in time with the dispersionless injection observed at RBSP-B. The dispersionless injection observed by RBSP-B and the sharp dipolarization observed by RBSP-B and ETS-8, both located at ∼ 21 h, indicate that both spacecraft are inside the current disruption region.
3. At 11:30 UT. A large-scale SCW is formed from ∼ 21 h to 2 h in MLT. The dispersed proton injection observed by RBSP-B and the lack of sharp dipolarization signatures in either RBSP-B or ETS-8 magnetic field data indicate that both spacecraft are now outside of the current disruption region, at the westward edge of it.

### 3.2. Inner Magnetosphere Conditions Measured by RBSP-B

Figure 9 shows plasma and magnetic field properties at the RBSP-B location throughout the interval. For the pressure, density, and temperature calculations we have used both HOPE (30 eV–45 keV) and RBSPICE (45 keV–600 keV) proton intensities, covering an extensive energy range, even though we only plot the RBSPICE intensities. We have also used the formula by *Wolf et al.* [2006] for the calculation of the flux tube volume $V = \int \frac{ds}{B}$ by single-point measurements. The formula provides a good estimate of the flux tube volume when $\sqrt{B_x^2 + B_y^2}/B_z \leq 3$ and the perpendicular flow < 150 km/s. The ratio $\sqrt{B_x^2 + B_y^2}/B_z$ is well below 1 throughout the interval of interest (Figure 9c), and the flow is also lower than the limit above (not shown). The flux tube volume combined with the plasma pressure is used to calculate the local entropy parameter $PV^{5/3}$. The entropy parameter (referred to as entropy hereafter) is critical to the earthward plasma transport since, as mentioned in section 1, theory suggests that flux tubes of lower entropy than the neighboring ones can move farther earthward due to interchange instability. The entropy calculation assumes isotropic pressure, which is the case for this interval (not shown).

As it can be seen in Figure 9, the dispersionless injection observed at RBSP-B at 11:21 UT has all the characteristics of a dipolarization front penetrating deep into the inner magnetosphere similar to those reported by *Runov et al.* [2011] in their multicase study of dipolarization fronts in the plasma sheet. These characteristics are (i) sharp dipolarization of the magnetic field, (ii) increase of the ion pressure due to density increase right ahead of the front, and (iii) decrease of the pressure and density and increase of temperature right behind the front. In Figure 9g we show flux tube volume in red and the entropy in black. Note that right before the dipolarization there is an increase in entropy and then a sharp decrease to lower than predipolarization values, both changes due to both flux tube volume and pressure increase ahead and decrease behind the front. The entropy reduction behind the front agrees with the theory of depleted, low-entropy flux tube penetrating in the inner magnetosphere. In fact, both entropy reduction and the decrease (increase) of low- (high-) energy particle intensities behind the front have been predicted by Rice Convection Model - Equilibrium (RCM-E) simulations of an idealized plasma bubble injection [*Yang et al.*, 2011]. The entropy recovers to predipolarization values around 11:30 UT when the lower energy proton intensities (40–67 keV) of the second, dispersed injection start increasing. Another sharp drop of the entropy occurs at ~ 11:43 UT, due to a sharp pressure decrease when the higher-energy proton intensities (above 81 keV) of the second injection drop to their preinjection values (the flux tube volume gradually decreases at that time).

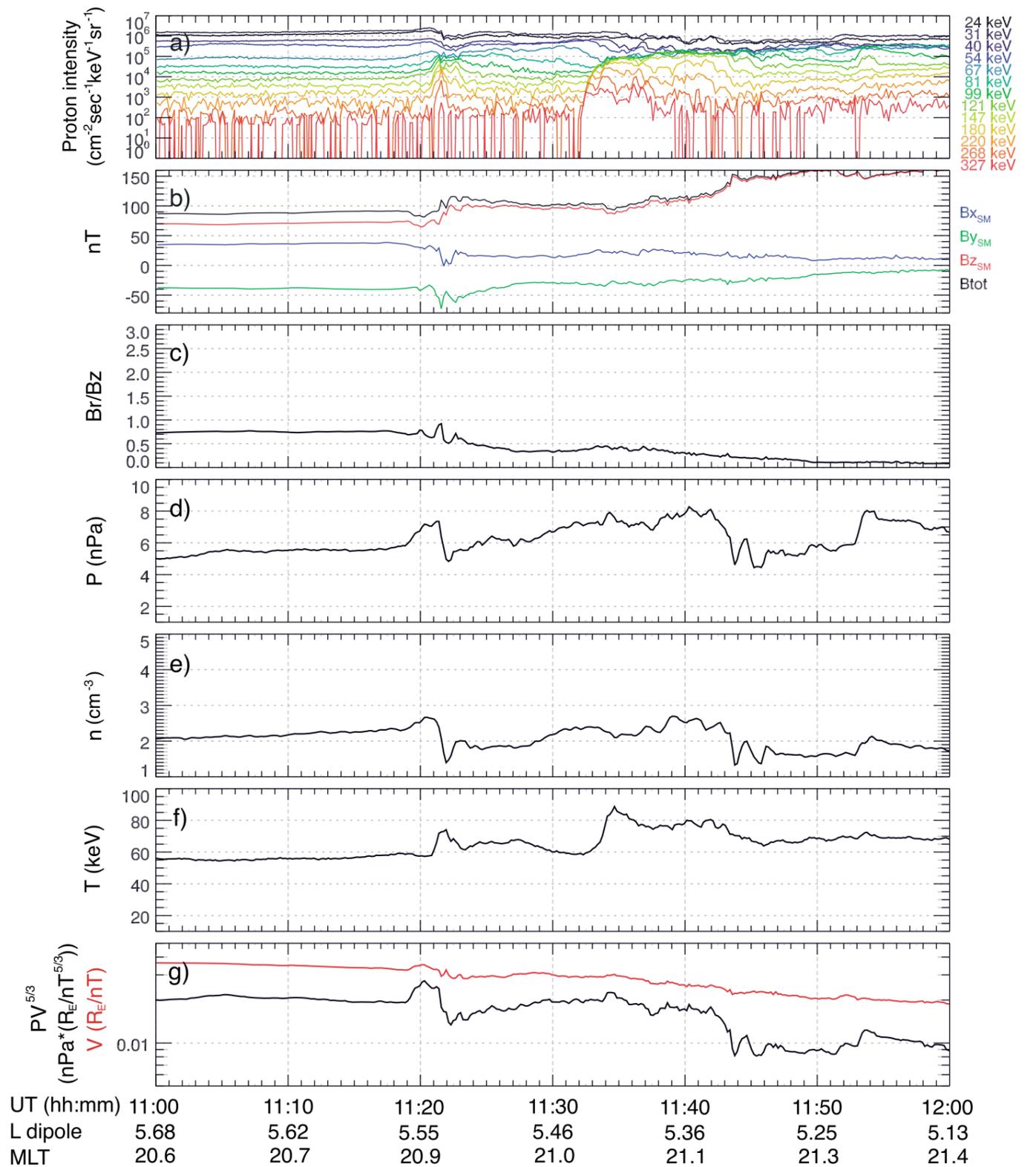

**Figure 9.** (a) RBSPICE proton intensities; (b) $B_x$, $B_y$, $B_z$, and $B_{tot}$; (c) $\sqrt{B_x^2 + B_y^2}/B_z$; (d) combined RBSPICE and HOPE proton pressure; (e) combined RBSPICE and HOPE number density; (f) combined RBSPICE and HOPE proton temperature; and (g) flux tube volume in red and entropy parameter in black.

These results indicate that the inner magnetosphere plasma conditions in the RBSP-B region change after each injection in a similar manner. Pressure and density increase before and decrease after the dipolarization, while the temperature increases after the dipolarization. However, there is a difference between the changes of the plasma properties that the two injections exhibit, and that is the timescale at which these changes occur; for the dispersionless injection changes occur within 3 min, while for the dispersed one the changes occur within more than 10 min. We discuss the implications of these timescales in the next section.

### 3.3. Implications of Injection Pulse Timescale in the Inner Magnetosphere

One of the questions posed in section 2.1 was whether such difference in the timescales of the injections observed is indicative of the spatial extent or temporal evolution of these injections. From ground magnetometer data we have established that a weak current wedge-like system of ~ 2 h MLT extent is formed around 11:20 UT associated with the dispersionless injection (RBSP-B is inside but close to the upward FAC of that current system at that time). The ground magnetic field perturbations associated with this current wedge cease to exist by 11:30 UT. Starting at ~ 11:30 UT, we have the intensification and westward expansion of the large-scale SCW with its upward FAC (westward edge) at ~21 h and its downward FAC (eastward edge) at ~ 2 h in MLT. At that time, RBSP-B is outside the SCW. As we discussed in section 3.1, although there is a prolonged dispersed proton injection observed at RBSP-B at ~ 11:32 UT, there is no clear prolonged dispersionless ion or electron injection observed at LANL-01A (located inside the SCW at that time) but only a short-timescale enhancement of proton intensities at ~ 11:29 UT and a gradual enhancement of the electron intensities starting at 11:30 UT. There are two possible scenarios that could explain this result: (i) Even though the SCW expands ~ 5 h in local time, the injection itself was more localized in azimuth and did not extend to midnight; that is, it extended less than 2 h in local time. In that case, the long duration of the injection pulse has to be attributed to a temporal component; that is, the injection is not instantaneous. (ii) Due to magnetic field variations, LANL-01A could be moving in and out of the center of the current sheet, causing short-timescale intensity enhancements and drops, rather than a prolonged in time injection. Let us examine the second scenario further: The azimuthal drift time for 90° pitch angles is $T_D = 43.8/(L \cdot E)$ [e.g., *Parks*, 2004], where $E$ is the energy in MeV and $T_D$ is time in minutes. Thus, 180 keV particles of 90° pitch angle at $L = 5.5$ have azimuthal drift time around the Earth $T_D = 44$ min. Therefore, their drift speed is $V_D = 2\pi \cdot 5.5 \cdot R_E/(T_D \cdot 60) \sim 83$ km/s. If the injection extent is 5 h in MLT, then the time they will need to travel from the eastward edge of the injection to the westward one is $(\pi \cdot 5.5 \cdot R_E \cdot 5/12)/83 \sim 555$ s $\sim 9.2$ min. However, the intensity of the 180 keV channel during the dispersed injection observed by RBSP-B remains elevated for ~12 min (11:32–11:44 UT). Thus, even in the second scenario, the assumption that the injection is not instantaneous is necessary in order to explain the duration of the injection. Similar conclusions regarding the temporal versus spatial component contributing to the duration of injection pulses were reported by *Reeves et al.* [1990].

We have showed that there is a temporal component (i.e., the source of the injected particles remains "on" for several minutes on the affected L shells) contributing to the duration of the second injection. We next examine whether that is the case for the first injection. Since the current wedge associated with the first injection extends ~2 h in MLT, then 180 keV particles need $(\pi \cdot 5.5 \cdot R_E \cdot 2/12)/83 \sim 222$ s $\sim 3.7$ min, which is very close to the duration of the 180 keV pulse (11:20–11:24 UT). Therefore, we can conclude that the first injection was instantaneous. Since clearly there is a difference in the temporal evolution of the two injections, the question arises as to whether the magnetic field reconfiguration during these two events plays a role in that temporal evolution.

Injections have been associated with sharp dipolarizations of the magnetic field. And indeed, that is the case for the dispersionless injection in this event study. However, as we have mentioned earlier in the paper, the dispersed injection is accompanied by a gradual dipolarization of the magnetic field. The magnetic field starts dipolarizing at RBSP-B at ~ 11:34:30 UT, approximately 2 min after the intensity of the highest energy of the dispersed injection starts rising and around the time that intensities of lower energy channels (40 keV–67 keV) drop. The fact that there is no sharp dipolarization associated with the dispersed injection could be due to RBSP-B not being located inside the current disruption region, as it is implied by the energy dispersion. In this case, the result depicts the caveats of estimating the MLT extent of an injection using only particle spectra. In fact, inside geosynchronous orbit, 1–2 min energy dispersion, an upper limit that has been extensively used to identify injections as dispersionless, could be too long. On the other hand, the gradual dipolarization observed by RBSP-B is a signature observed by the rest of the satellites as well and evidently is propagating earthward, as we already discussed in section 2.2: it is first seen by TH-D at ~11:25 UT, then by GOES 15 and ETS-8 at ~11:33 UT, and finally by RBSP-B at 11:34:30 UT. We should point out that GOES 15 also sees fluctuations of the magnetic field that correspond to the Pi2 pulsations observed by midnight and postmidnight stations. With TH-D not being in the neutral sheet for large portion of the event, and with no other satellites providing magnetic field data in the region between 21 h and 2 h in MLT in the inner magnetosphere, we cannot conclusively determine whether there was a sharp dipolarization in that region that none of the other spacecraft saw due to their location or whether a larger-scale gradual dipolarization in the inner magnetosphere is responsible for the large duration of the dispersed injection observed by RBSP-B.

## 4. Summary and Conclusions

We have presented here a case study of a substorm event that occurred during a small storm ($Dst \sim -40$ nT) on 14 July 2013. Two energetic particle injections were observed by RBSP-B deep inside geosynchronous orbit, only 10 min apart yet exhibiting different dipolarization signatures as well as duration. The first event is a dispersionless, short-timescale (~3 min) energetic proton and electron injection accompanied by sharp dipolarization of the magnetic field. The second event is a dispersed energetic proton injection exhibiting ~1.5 min energy dispersion, lasting ~10 min, and is accompanied by a gradual dipolarization of the magnetic field. Motivated by this observation, we combined ground magnetometer data and in situ particle and magnetic field data from geosynchronous satellites GOES 15, LANL-01A, and ETS-8 and TH-D downtail in order to investigate the spatial extent of these injections, their temporal evolution, and their effects in the inner magnetosphere plasma properties. We found that the first one is associated with the formation of a weak current system possibly caused by a localized low-entropy bubble extending ~ 2 h in MLT. According to energy-dependent drift calculations, the duration of the pulse of the injection can be attributed to the time that protons take to travel from the eastward to the westward edge of that bubble. Following similar arguments, we found that the second one extends no more than 5 h in MLT, is associated with the development of a large-scale substorm current wedge, and unlike the first one has a temporal component contributing to the duration of the injection pulse. However, we cannot conclusively determine whether this is related to the gradual dipolarization observed by both RBSP-B and all other spacecraft, due to a lack of magnetic field measurements between 21 and 2 h in MLT. Van Allen Probes' high-energy and high time resolution data with unprecedented energy coverage in the inner magnetosphere allowed us to calculate the changes in plasma properties such as pressure, number density, temperature, and the entropy parameter during both injections. The first one has all the characteristics of a dipolarization front, typically observed in the near-Earth plasma sheet, penetrating deep inside geosynchronous orbit: sharp dipolarization of the magnetic field, increase of the pressure and density right ahead of the front, decrease of the pressure and density and increase of temperature right behind the front, and sharp decrease of the entropy to lower than predipolarization values, all being consistent with the theory of a depleted, low-entropy flux tube penetrating into the inner magnetosphere. The second one, even though we cannot determine whether or not is associated with a sharp dipolarization of the magnetic field, exhibits similar changes in the plasma properties but over longer timescales. Interestingly, recent results have interpreted the formation of SCW (which in our event coincides with the second, long-timescale injection) as the result of multiple flow bursts associated with low-entropy bubbles penetrating into the inner magnetosphere (similar to the first, localized, short-timescale injection of our event) [Lyons et al., 2012; Sergeev et al., 2014].

Our results suggest that there are different spatial and temporal scales at which injections can occur in the inner magnetosphere and depict the necessity of multipoint observations of both particle and magnetic field data in order to determine these scales.

### Acknowledgments

The authors thank team discussions with the larger RBSPICE and Van Allen Probes teams. The RBSPICE instrument was supported by JHU/APL subcontract 937836 to the New Jersey Institute of Technology under NASA prime contract NAS5-01072. MG was also supported by NSF grant AGS-1303646 and the International Space Science Institute's (ISSI) International Teams program. SO was supported by NASA grant NNX12AJ52G. D.L.T. was supported by NASA's THEMIS (contract NAS5-02099) and Van Allen Probes (contract NAS5-01072) missions. J.V.R. was supported by National Geophysical Data Center (NGDC) Task II under the CIRES Cooperative Agreement between NOAA and the University of Colorado. MN was supported by the Ministry of Education, Culture, Sports, Science and Technology (MEXT) Grant-in-Aid Scientific Research (B) (grant 25287127). Solar wind data and $AL$ and $SYM$-$H$ indices were retrieved from OMNIweb service. The $Wp$ index was retrieved from http://s-cubed.info/. One second resolution magnetometer data were collected at various magnetic observatories within the INTERMAGNET project. One minute resolution magnetometer data were collected at various magnetic observatories within the SuperMAG project. We thank the national institutes that support the magnetic observatories and INTERMAGNET (www.intermagnet.org) and SuperMAG (http://supermag.jhuapl.edu/) projects for promoting high standards of magnetic observatory practice. Magnetic field data from the geosynchronous ETS-8 satellite were provided by the Japan Aerospace Exploration Agency upon request. THEMIS-D data were retrieved with the Space Physics Environment Data Analysis System (SPEDAS) software (http://themis.igpp.ucla.edu/software.shtml). GOES 15 magnetometer data were retrieved from CDAweb service. GOES MAGPD data are available upon request from NGDC Van Allen Probes RBSPICE data were retrieved from http://rbspice.ftecs.com/, and HOPE and MagEIS data were retrieved from http://www.rbsp-ect.lanl.gov/rbsp_ect.php. LANL-01A data were provided by the Los Alamos National Laboratory upon request.

Michael Liemohn thanks the reviewers for their assistance in evaluating this paper.